\documentclass[12pt]{article}

\usepackage{amsmath}
\usepackage{amssymb}
\usepackage[bookmarks=false]{hyperref}

\newcommand{\be}{\begin{equation}}
\newcommand{\ee}{\end{equation}}

\newcommand{\ui}{{\underline{i}}}
\newcommand{\uj}{{\underline{j}}}

\newcommand{\tr}{{\rm tr}\,}

\newcommand{\cN}{{\cal N}}

\newcommand{\bea}{\begin{eqnarray}}
\newcommand{\eea}{\end{eqnarray}}
\newcommand{\ba}{\begin{array}}
\newcommand{\ea}{\end{array}}
\newcommand{\nn}{\nonumber}
\newcommand{\half}{\frac{1}{2}}
\newcommand{\qter}{\frac{1}{4}}
\newcommand{\mc}{\mathcal}

\newcommand{\p}{\partial}

\newcommand{\pr}{\prime}
\newcommand{\ov}{\overline}

\newcommand{\wt}{\widetilde}

\newcommand{\eps}{\varepsilon}

\newcommand{\al}{\alpha}

\newcommand{\da}{\delta}
\newcommand{\om}{\omega}
\newcommand{\Ga}{\Gamma}
\newcommand{\ga}{\gamma}
\newcommand{\La}{\Lambda}

\newcommand{\si}{\sigma}
\newcommand{\ta}{\theta}
\newcommand{\Om}{\Omega}

\begin{document}

\thispagestyle{empty}

\vspace{1cm}

\begin{center}
{\Large\bf Wess-Zumino term in the \\[3mm]
$\cN=4$ SYM effective action revisited} \vspace{1.5cm}

{\large\bf Dmitry~V.~Belyaev\,{}$^\dag$ {\it and}\hspace{5pt}
 Igor~B.~Samsonov$\,{}^{\star}$\footnote{On leave from
Tomsk Polytechnic University, 634050 Tomsk, Russia}
\\[8pt]
\it\small $^\dag$ Institute for Fundamental Theory, Department of Physics,
\\ University of Florida, Gainesville, FL 32611, USA\\
{\tt email:\ belyaev@phys.ufl.edu}\\[8pt]
$^\star$INFN, Sezione di Padova, 35131 Padova, Italy\\
{\tt email:\ samsonov@mph.phtd.tpu.ru}}
\end{center}
\vspace{0.5cm}

\begin{abstract}

The low-energy effective action for the $\cN=4$ super Yang-Mills on the Coulomb branch is known to include an $SO(6)$-invariant Wess-Zumino (WZ) term for the six scalar fields. For each maximal, non-anomalous subgroup of the $SU(4)$ R-symmetry, we find a four-dimensional form of the WZ term with this subgroup being manifest. We then show that a recently proposed expression for the four-derivative part of the effective action in $\cN=4$ $USp(4)$ harmonic superspace yields the WZ term with manifest $SO(5)$ R-symmetry subgroup. The $\cN=2$ $SU(2)$ harmonic superspace form of the effective action produces the WZ term with manifest $SO(4)\times SO(2)$. We argue that there is no four-dimensional form of the WZ term with manifest $SU(3)$ R-symmetry, which is relevant for $\cN=1$ and $\cN=3$ superspace formulations of the effective action.

\end{abstract}

\newpage
\tableofcontents

\numberwithin{equation}{section}

\section{Introduction}

Wess-Zumino (WZ) terms arise in low-energy quantum effective actions as a consequence of anomalies in global symmetries \cite{Wess:1971yu,Witten:1983tw}. In a four-dimensional gauge theory, with gauge group $G_\text{g}$ and global symmetry group $G_\text{gl}$, the anomaly in $G_\text{gl}$ can appear in a `global-gauge-gauge' or in a `global-global-global' triangle diagram. In the former case, the global symmetry is broken at the quantum level: the Noether current of $G_\text{gl}$ is not conserved and the quantum effective action has non-zero variation under $G_\text{gl}$. However, if only the `global-global-global' diagram is anomalous, $G_\text{gl}$ is \emph{not} broken at the quantum level: the current is conserved and the effective action is invariant. Still, the anomaly manifests itself in the appearance of the WZ term, which can be understood using 't Hooft anomaly matching argument \cite{'tHooft:1979bh,Weinberg:1996kr}.

This is precisely what happens in the $\cN=4$ super Yang-Mills (SYM) theory \cite{Brink:1976bc,Gliozzi:1976qd}, which has global $SU(4)$ R-symmetry with anomalous `global-global-global' diagram.~\footnote
{
The $\cN=4$ SYM is a \emph{chiral} theory. Its fermionic degrees of freedom are positive helicity {\bf 4} and negative helicity $\bf\bar{4}$ of $SU(4)_R$, all in the adjoint of the gauge group $G_\text{g}$. The $SU(4)_R$ currents are chiral, and there is an anomaly in their triangle diagram proportional to the number of charged fermions, and thus to the dimension $|G_\text{g}|$ of the gauge group \cite{Sohnius:1981sn,Witten:1998qj,Freedman:1998tz,Chalmers:1998xr,Aharony:1999ti}. There are, however, no anomalies involving the gauge group currents, and the full superconformal group $PSU(2,2|4)$ is unbroken at the quantum level \cite{White:1992wu}.
}
When the gauge group $G_\text{g}$ is spontaneously broken to a subgroup $H_\text{g}$, and $|G_\text{g}|-|H_\text{g}|$ massive gauginos are integrated out, a WZ term \cite{TZ} appears in the effective action with the coefficient proportional to $|G_\text{g}|-|H_\text{g}|$ so that 't Hooft anomaly matching is satisfied \cite{Weinberg:1996kr,Intriligator:2000eq}. As the scalars receiving vacuum expectation values are in the \emph{adjoint} of $G_\text{g}$, the unbroken $H_\text{g}$ necessarily has a $U(1)$ subgroup \cite{Slansky:1981yr}, so that the theory is `on the Coulomb branch.' 

The basic example has $G_\text{g}=SU(2)$ spontaneously broken to
$H_\text{g}=U(1)$ \cite{Fayet:1978ig}. The massless degrees of
freedom then constitute one $\cN=4$ abelian multiplet containing
the Maxwell field-strength $F_{m n}$, six scalars $X_A$ and
four gauginos (in the {\bf 1}, {\bf 6} and {\bf 4} representation
of $SU(4)_R$, respectively). The low-energy effective action, $\Ga$,
is conformal and $\cN=4$ supersymmetric \cite{Seiberg:1988ur},
albeit very non-local. Within the derivative expansion of the
effective action \cite{Henningson:1995eh}, the two-derivative part, $\Ga_2$, is given by the classical $\cN=4$ super Maxwell action \cite{Seiberg:1994rs}. The first non-trivial contribution is given by the four-derivative part, $\Ga_4$, which includes the so-called `$F^4/X^4$' term \cite{DS,S16},
\bea
\label{F4X4}
\frac{1}{(8\pi)^2}\int d^4x \frac{1}{(X_A X_A)^2}\Big(F_{m n}F^{n k}F_{k l}F^{l m}-\qter(F_{p q}F^{p q})^2\Big) \
,
\eea
as well as the pure-scalar WZ term \cite{TZ,Intriligator:2000eq} (shown here in its five-dimensional form),
\bea
\label{WZterm}
-\frac{1}{60\pi^2}\int d^5x\,\eps^{M N K L P}\eps^{A B C D E F} \,
\frac{1}{|X|^6}
X_A \p_M X_B \p_N X_C \p_K X_D \p_L X_E \p_P X_F \ ,
\eea
where $|X|^2=X_A X_A$. In this paper, we will analyze superfield actions which include both (\ref{F4X4}) and (\ref{WZterm}) among their components and thus represent the $\cN=4$ SYM low-energy effective action $\Gamma_4$.

To write the WZ term (\ref{WZterm}) directly as a $d=4$ integral, one has to
sacrifice part of the manifest $SO(6)$ R-symmetry. The
't Hooft anomaly matching argument \cite{'tHooft:1979bh,Weinberg:1996kr}
tells us that all anomalous R-symmetry generators must transform the
four-dimensional WZ term into a total divergence, and therefore
\emph{anomalous R-symmetry subgroups cannot remain manifest.} Among the
four maximal subgroups of $SO(6)_R\simeq SU(4)_R$ \cite{Slansky:1981yr,PRbook}
(defined by the decomposition of the {\bf 4} of $SU(4)_R$),
\bea
\label{msubs}
\ba[b]{rclcrcl}
&& SU(3)\times U(1), &\quad& {\bf 4} &=& {\bf 3}_{+1}+{\bf 1}_{-3} \\
SO(5) &\simeq& USp(4), &\quad& {\bf 4} &=& {\bf 4} \\
SO(4)\times SO(2) &\simeq& SU(2)\times SU(2)\times U(1), &\quad&
{\bf 4} &=& ({\bf 2},{\bf 1})_{+1}+({\bf 1},{\bf 2})_{-1} \\
SO(3)\times SO(3) &\simeq& SU(2)\times SU(2), &\quad& {\bf 4} &=& ({\bf 2},{\bf 2})\ ,
\ea
\eea
the first subgroup is anomalous, whereas the other three are non-anomalous.~\footnote{The anomaly is absent for the $USp(4)$ and $SU(2)\times SU(2)$ subgroups because the {\bf 4} of $USp(4)$ and {\bf 2} of $SU(2)$ are self-conjugate \cite{GGanomal}. The potential $U(1)$ anomaly for the $SU(2)\times SU(2)\times U(1)$ subgroup cancels due to the symmetric $U(1)$ charge assignments in ${\bf 4}= ({\bf 2},{\bf 1})_{+1}+({\bf 1},{\bf 2})_{-1}$.}
For each of the non-anomalous subgroups, we will explicitly construct a four-dimensional form of the WZ term with this subgroup being manifest. For the $SU(3)\times U(1)$ subgroup, this is not possible, according to the above argument.

Supersymmetry of the $\cN=4$ SYM effective action $\Ga$ would be best utilized by writing it as a superspace functional.
In this paper, we will consider the superspace form of $\Ga$ \emph{on classical shell}. Namely, we will constrain the fields of the $\cN=4$ gauge multiplet to satisfy their equations of motion,
\bea
\label{oncshell}
\p^m F_{m n}=\Box X_A=0 \ .
\eea
As is well-known, the effective action $\Ga$ restricted to on-shell fields is related to the S-matrix functional \cite{FadSlav}. 

On shell, the four-derivative part of the effective action, $\Ga_4$, turns out to be simplest in $\cN=4$ $USp(4)$
harmonic superspace \cite{IKVO,BLS}. The gauge multiplet is then represented
by a single $\cN=4$ superfield strength $\mc{W}$, and we find that
\bea
\label{BLS}
\Ga_4=-\frac{1}{96\pi^2}\int d\zeta dv \ln\frac{\mc{W}}{\La} \ ,
\eea
where the integral is over an analytic subspace of the $\cN=4$ superspace. This form of the $\cN=4$ SYM effective action was suggested in \cite{BLS}. We will confirm its correctness by demonstrating that it does reproduce both the `$F^4/X^4$' and the WZ term.

In the more familiar $\cN=2$ $SU(2)$ harmonic superspace \cite{GIKOS,book}, the $\cN=4$ gauge multiplet is described by one $\cN=2$ gauge superfield strength $W$ and one $\cN=2$ hypermultiplet superfield $q_a^{+}$. The expression for $\Ga_4$ in this superspace was found by Buchbinder and Ivanov \cite{BuIv} to have the following form~\footnote{
The hypermultiplet-independent part of (\ref{DS+BUIV}) is given by the non-holomorphic potential \cite{Henningson:1995eh} found in \cite{deWit:1996kc,DS}. The overall coefficient has been calculated perturbatively in \cite{non-hol1,non-hol2,non-hol3,non-hol4,non-hol5} and is directly related to the coefficient in (\ref{F4X4}).
}
\bea
\label{DS+BUIV}
\Ga_4=\frac{1}{(4\pi)^2}\int d^4x d^{8}\ta du \left\{
\ln\frac{W}{\La}\ln\frac{\bar W}{\La}
+\sum_{n=0}^\infty\frac{1}{n^2(n+1)}\left(-\frac{q^{+ a}q_a^{-}}{W\bar W}\right)^n
\right\} \ ,
\eea
where the integration is over the full $\cN=2$ harmonic superspace. It has been known that this action contains the `$F^4/X^4$' term (\ref{F4X4}). We will demonstrate that it contains the WZ term (\ref{WZterm}) as well. The actions (\ref{BLS}) and (\ref{DS+BUIV}) are, therefore, equivalent on classical shell.

This paper is organized as follows. In Section \ref{sec-WZ}, we will present the manifestly $SO(5)$, $SO(4)\times SO(2)$ and $SO(3)\times SO(3)$-invariant forms of the $SO(6)$-invariant WZ term (\ref{WZterm}). These will follow from a more general construction valid in any dimension.
In Sections \ref{sec-EAN4} and \ref{sec-EAN2}, we will motivate and analyze the superspace actions (\ref{BLS}) and (\ref{DS+BUIV}), respectively, giving details for the corresponding superspaces in Appendices \ref{HSS4} and \ref{HSS2}. In Section \ref{sec-summary}, we will summarize our results and discuss some related questions.

\section{$SO(6)$-invariant WZ term}
\label{sec-WZ}

The WZ term (\ref{WZterm}) corresponds to a particular case of the $d$-dimensional WZ term in the $SO(d+2)$ sigma model \cite{Patani:1976gh,Braaten:1985is,Abanov:1999qz}. In this section, we will construct $d$-dimensional Lagrangian forms for this term with manifest $SO(n)\times SO(d+2-n)$ subgroups of $SO(d+2)$, and then specialize to $d=4$. For $n=d+1$, our results reproduce those in \cite{Braaten:1985is}. To the best of our knowledge, the results for other $n$ are new.

\subsection{Various forms of the $SO(d+2)$-invariant WZ term}

The $SO(d+2)$-invariant $d$-dimensional WZ term is constructed out of Goldstone bosons parametrizing $S^{d+1}=SO(d+2)/SO(d+1)$. Introducing a $(d+2)$-component scalar field $X_A$, with $A=1,\dots,d+2$, we will use the normalized scalar field $Y_A$,
\bea
Y_A=\frac{X_A}{|X|}, \quad |X|=\sqrt{X_A X_A}, \quad Y_A Y_A=1 \ ,
\eea
to parametrize $S^{d+1}$. The WZ term is given by the $(d+1)$-dimensional integral of a $(d+1)$-form that generates the de Rham cohomology group $H^{d+1}(S^{d+1};\mathbb{R})=\mathbb{Z}$ \cite{dHWein}. The simplest choice of such a form corresponds to the volume form on $S^{d+1}$,
\bea
\om_{d+1} &=& \frac{\eps^{A_1\dots A_{d+2}}}{(d+1)!}Y_{A_1}dY_{A_2}\wedge dY_{A_3}\wedge\dots\wedge dY_{A_{d+2}} \nn\\
&=& d^{d+1}x\frac{\eps^{A_1\dots A_{d+2}}}{(d+1)!}\,
\eps^{M_1\dots M_{d+1}}Y_{A_1}\p_{M_1}Y_{A_2}\dots\p_{M_{d+1}}Y_{A_{d+2}} \ .
\eea
The conventionally normalized WZ term is then defined as follows \cite{Braaten:1985is,Abanov:1999qz}
\bea
\label{dWZ}
S_{WZ}^{(d)}=-N\frac{(d/2)!}{\pi^{d/2}}\int_{\Om_Y} \om_{d+1} \ .
\eea
Here $\Om_Y$ is a hemisphere in $S^{d+1}$ whose boundary, $\p\Om_Y$, is the image of the $d$-dimensional space-time, viewed as a large $S^d$, under the map $Y_A(x)$ \cite{Witten:1983tw,Gaume}.
For any integer $N$, choosing another hemisphere changes $S_{WZ}^{(d)}$ by $2\pi$ times an integer.

Let us now split the index $A$ into $a=1,\dots,n$ and $a^\pr=n+1,\dots n+m$, where we defined $m=d+2-n$. With the convention $\eps^{1\dots(n+m)}=\eps^{1\dots n}\eps^{n+1\dots n+m}$, we then have
\bea
\om_{d+1}=\frac{1}{m}\om_{n-1}\wedge d\om_{m-1}^\pr
+(-)^n\frac{1}{n}d\om_{n-1}\wedge\om_{m-1}^\pr \ ,
\eea
where
\bea
\om_{n-1} = \frac{\eps^{a_1\dots a_n}}{(n-1)!}Y_{a_1}dY_{a_2}\dots dY_{a_n}, \quad
\om_{m-1}^\pr = \frac{\eps^{a_1^\pr\dots a_m^\pr}}{(m-1)!}Y_{a_1^\pr}dY_{a_2^\pr}\dots dY_{a_m^\pr} \ .
\eea
Introducing $y=Y_a Y_a=1-Y_{a^\pr}Y_{a^\pr}$, we find the following useful identities
\bea
\label{dyom}
dy\wedge\om_{n-1}=\frac{2}{n}y d\om_{n-1}, \quad
dy\wedge\om_{m-1}=-\frac{2}{m}(1-y) d\om_{m-1} \ ,
\eea
where we used that the antisymmetrization in $(n+1)$ $n$-dimensional indices yields zero. It then follows that
\bea
\label{dyomom}
\om_{d+1}=(-)^{n}\frac{dy\wedge\om_{n-1}\wedge\om_{m-1}^\pr}{2y(1-y)} \ .
\eea
Next, we take the following ansatz~\footnote{
The volume form $\om_{d+1}$ is closed, but not exact. This is consistent with (\ref{omf}) only if $f(y)$ is singular at some value of $y$ in the interval $0\leq y\leq 1$.
}
\bea
\label{omf}
\om_{d+1}=d\Big(f(y)\om_{n-1}\wedge\om_{m-1}^\pr\Big) \ ,
\eea
and also bring it to the form (\ref{dyomom}) using the identities (\ref{dyom}). We then immediately find that $f(y)$ must satisfy the following differential equation,
\bea
\label{feq}
\frac{d}{dy}f(y)+\frac{1}{2}\left(\frac{n}{y}-\frac{m}{1-y}\right)f(y)=\frac{(-)^{n}}{2y(1-y)} \ .
\eea
Its general solution is given by~\footnote{
$B(n,m)=\Ga(n)\Ga(m)/\Ga(n+m)$ is the Euler beta function, and
$B_y(n,m)=\int_0^y dt\, t^{n-1}(1-t)^{m-1}$ is the incomplete beta function satisfying $B_1(n,m)=B(n,m)$.
}
\bea
\label{fnm}
f(y)=\frac{(-)^n}{2y^{n/2}(1-y)^{m/2}}
\left\{B_y\left(\frac{n}{2},\frac{m}{2}\right)-C\, B\left(\frac{n}{2},\frac{m}{2}\right)\right\} \ ,
\eea
where $C$ is the constant of integration. The solution is regular at $y=0$ if $C=0$ and regular at $y=1$ if $C=1$.
Choosing $f(y)$ that is non-singular in $\Omega_Y$ and using Stokes's theorem, we obtain the $d$-dimensional form of the WZ term with manifest $SO(n)\times SO(m)$ invariance, ($d=n+m-2$),
\bea
\label{WZnm}
S_{WZ}^{(d)}=-N\frac{(d/2)!}{\pi^{d/2}}
\frac{\eps^{a_1\dots a_n}}{(n-1)!}
\frac{\eps^{a_1^\pr\dots a_m^\pr}}{(m-1)!}
\int_{\p\Om_Y}\!\! f(Y_a Y_a)
Y_{a_1}dY_{a_2}\dots dY_{a_n}
Y_{a_1^\pr}dY_{a_2^\pr}\dots dY_{a_m^\pr} \, .
\eea
The residual transformations from $SO(d+2)$ vary the integrand in this expression into an exact $d$-form, consistent with the fact that $S_{WZ}^{(d)}$ is $SO(d+2)$ invariant.

\subsection{$SO(6)$ WZ term with manifest $SO(5)$}

For $d=4$, (\ref{dWZ}) gives the manifestly $SO(6)$-invariant WZ term
\bea
\label{WZso6}
S_{WZ}^{(4)}=-\frac{N}{60\pi^2}\int_{\Om_Y} \eps^{A B C D E F}Y_A dY_B\wedge dY_C\wedge dY_D\wedge dY_E\wedge dY_F \ ,
\eea
which reduces to (\ref{WZterm}) for $N=1$. Using (\ref{WZnm}) with $n=5$ and $m=1$, we then obtain the 4-dimensional form of this WZ term with manifest $SO(5)$ invariance,
\bea
\label{WZso5}
S_{WZ}^{(4)} &=& \frac{N}{60\pi^2}\int_{\p\Om_Y}\eps^{abcde}\;\frac{g(z)}{Y_6^5}
Y_a dY_b\wedge dY_c\wedge dY_d\wedge dY_e \nn\\
&=& \frac{N}{60\pi^2}\int d^4x\,\eps^{mnpq}\eps^{abcde}\;\frac{g(z)}{X_6^5}X_a\p_m X_b\p_n X_c\p_p X_d\p_q X_e \ ,
\eea
where $m=0,1,2,3$ is the four-dimensional space-time index, $a=1,2,3,4,5$ is the $SO(5)$ index, and we defined $g(z)=-5(1-y)^3 f(y)$ with
\bea
z^2=\frac{y}{1-y}=\frac{Y_a Y_a}{Y_6^2}=\frac{X_a X_a}{X_6^2} \ .
\eea
This function satisfies the following equation
\bea
z\frac{d}{dz}g(z)+5g(z)=\frac{5}{(1+z^2)^3} \ ,
\eea
and its solution that is regular at $z=0$, with $g(0)=1$, is given by
\bea
\label{gsum}
g(z)=\frac{5}{8z^5}\left(3\arctan{z}-\frac{z(3+5z^2)}{(1+z^2)^2}\right)
=\frac{5}{2}\sum_{n=0}^\infty\frac{(n+2)(n+1)}{2n+5}(-z^2)^n \ .
\eea
In Section \ref{sec-EAN4}, we will show that this function arises naturally from (\ref{BLS}), i.e. from
the $\cN=4$ SYM effective action in the $\cN=4$ $USp(4)$ harmonic superspace.

\subsection{$SO(6)$ WZ term with manifest $SO(4)\times SO(2)$}

When $n=4$ and $m=2$, the solution to (\ref{feq}) that is regular at $y=0$ is simply
\bea
f(y)=\frac{1}{4(1-y)} \ .
\eea
The form of the WZ term (\ref{WZso6}) with manifest $SO(4)\times SO(2)$ invariance is then
\bea
S_{WZ}^{(4)} &=& -\frac{N}{12\pi^2}\int_{\p\Om_Y}\eps^{abcd}\eps^{a^\pr b^\pr}
Y_a dY_b\wedge dY_c\wedge dY_d\wedge \frac{Y_{a^\pr} dY_{b^\pr}}{Y_{c^\pr} Y_{c^\pr}} \nn\\
&=& -\frac{N}{12\pi^2}\int d^4x\, \eps^{m n p q}\eps^{a b c d}\eps^{a^\pr b^\pr}
\frac{X_a\p_m X_b\p_n X_c\p_p X_d}{(X_e X_e+X_{d^\pr} X_{d^\pr})^2}
\frac{X_{a^\pr}\p_q X_{b^\pr}}{X_{c^\pr} X_{c^\pr}} \ ,
\eea
where now $a=1,2,3,4$ is the $SO(4)$ index, $a^\pr=5,6$ is the $SO(2)$ index, and we used that $1-y=Y_{c^\pr}Y_{c^\pr}$.
Introducing the following polar decomposition
\bea
X_6+i X_5=X e^{i\al} \ ,
\eea
we then find that
\bea
\label{WZso4so2}
S_{WZ}^{(4)}=\frac{N}{12\pi^2}\int d^4x\,\eps^{m n p q}\eps^{abcd}
\frac{X_a\p_m X_b\p_n X_c\p_p X_d}{(X_e X_e+X^2)^2}
\p_q\al \ .
\eea
In this form, the $SO(2)$ acts by shifting $\alpha$ by a constant. In Section \ref{sec-EAN2}, we will see this form of the WZ term arising from (\ref{DS+BUIV}), i.e. from the $\cN=4$ SYM effective action in the $\cN=2$ $SU(2)$ harmonic superspace.

\subsection{$SO(6)$ WZ term with manifest $SO(3)\times SO(3)$}

Using (\ref{WZnm}) with $n=3$ and $m=3$, we obtain the form of the WZ term (\ref{WZso6}) with manifest $SO(3)\times SO(3)$ invariance,
\bea
S_{WZ}^{(4)} &=& -\frac{N}{2\pi^2}\int_{\p\Om_Y}\eps^{abc}\eps^{a^\pr b^\pr c^\pr}f(y)
Y_a dY_b\wedge dY_c\wedge Y_{a^\pr}\, dY_{b^\pr}\wedge dY_{c^\pr} \ ,
\eea
where $y=Y_a Y_a=1-Y_{a^\pr}Y_{a^\pr}$. The solution for $f(y)$ is as given by (\ref{fnm}) with $C=0$. However, we will not discuss this form of the WZ term further in this work.

To summarize, we have found that the WZ term (\ref{WZterm}) can be written in three different four-dimensional forms, with manifest $SO(5)$, $SO(4)\times SO(2)$, or $SO(3)\times SO(3)$, respectively. The residual $SO(6)$ transformations are non-manifest symmetries of $S_{WZ}^{(4)}$: they vary the integrands in the corresponding expressions into total divergences. Conformal $SO(4,2)$ transformations similarly split into manifest (leaving the integrands invariant) and non-manifest (changing the integrands by total divergences) symmetries.

In the following sections, we will find that the $SO(5)$ and $SO(4)\times SO(2)$ forms of the WZ term correspond, respectively, to the $\cN=4$ and $\cN=2$ harmonic superspace formulations of the $\cN=4$ SYM effective action.

\section{Effective action in $\cN=4$ harmonic superspace}
\label{sec-EAN4}

In this section, we will establish the $\cN=4$ harmonic superspace form for the four-derivative part in the $\cN=4$ SYM effective action on the Coulomb branch. An extensive list of $\cN=4$ harmonic superspaces, with harmonics defined on various coset manifolds $G/H$, is given in \cite{IKVO}. We will use the one with $USp(4)/[U(1)\times U(1)]$ harmonics \cite{BLS}. This choice is motivated by the fact that the WZ term, which must be present in the effective action, can be chosen to respect manifest $USp(4)\simeq SO(5)$. Our conventions and basic features of the $\cN=4$ $USp(4)$ harmonic superspace are explained in Appendix~\ref{HSS4}.

\subsection{Scale-invariant effective action}

In $\cN=4$ $USp(4)$ harmonic superspace, the $\cN=4$ gauge multiplet is described by a constrained superfield $\mc{W}$. The constraints make the component fields satisfy on-shell equations (\ref{oncshell}).
In the bosonic sector, the component decomposition of $\cal W$ reads~\footnote{
The missing fermionic components are given explicitly in eq.~(5.21) of \cite{BLS}. They are naturally written using $USp(4)$ harmonics $u$. For the bosonic components, however, we found it more convenient to use the $SO(5)$ harmonics $v$ \cite{Z1,Z2}. The $v$'s are given in terms of $u$'s in (\ref{cor-harm}).
}
\bea
{\cal W}&=&\varphi+i X_a v^5_a
+\frac1{\sqrt2} (\theta^{(+,0)}_\alpha\theta^{(-,0)}_\beta
\sigma^{m\alpha}{}_{\dot\alpha}\sigma^{n\beta\dot\alpha}
-\bar\theta^{(0,+)}_{\dot\alpha}\bar\theta^{(0,-)}_{\dot\beta}
\sigma^{m\dot\alpha}{}_\alpha \sigma^{n\alpha\dot\beta})F_{mn}
\nn\\
&& -2i\theta^{(+,0)}_\alpha\bar\theta^{(0,+)}_{\dot\alpha}
\partial^{\alpha\dot\alpha}X_a (v^1_a-iv^2_a)
+2i\theta^{(-,0)}_\alpha\bar\theta^{(0,-)}_{\dot\alpha}
\partial^{\alpha\dot\alpha}X_a (v^1_a+iv^2_a)
\nn\\[3pt]
&& +2i\theta^{(+,0)}_\alpha\bar\theta^{(0,-)}_{\dot\alpha}
\partial^{\alpha\dot\alpha}X_a (v^4_a-iv^3_a)
+2i\theta^{(-,0)}_\alpha\bar\theta^{(0,+)}_{\dot\alpha}
\partial^{\alpha\dot\alpha}X_a (v^4_a+iv^3_a)
\nn\\[3pt]
&& +4\theta^{(+,0)}_\alpha\theta^{(-,0)}_\beta
\bar\theta^{(0,+)}_{\dot\alpha}\bar\theta^{(0,-)}_{\dot\beta}
\partial^{\alpha\dot\alpha}\partial^{\beta\dot\beta}
[\varphi-iX_a v^5_a]\ .
\label{x15.2}
\eea
Here $\varphi$ and $X_a$ are the six scalars split into {\bf 1} and {\bf 5} of $USp(4)$.
The $SO(5)$ harmonics $v_a^b$ and Grassmann coordinates $\ta$'s are defined in Appendix \ref{HSS4}.

In general, the effective action is a functional which can be written as a
superspace integral of some function of $\cal W$ and its covariant
superspace derivatives. We point out that the analytic measure
$d\zeta$, written explicitly in (\ref{meas},\ref{N4measure}),
yields eight Grassmann derivatives, or, equivalently, four
space-times ones. Hence, the four-derivative part of the effective action
$\Gamma_4$ is given by
\be
\Ga_4=\int d\zeta dv\, {\cal H}({\cal W})\,,
\label{S4}
\ee
with some function ${\cal H}({\cal W})$ of the superfield strength
without derivatives.
This function can be fixed using scale invariance of the
$\cN=4$ SYM effective action.
As the measure $d\zeta dv$ is dimensionless, ${\cal H}({\cal W})$ should
also be dimensionless. As $\mc{W}$ has mass dimension one, we have to
introduce a parameter $\La$ such that ${\cal W}/\Lambda$ is
dimensionless and choose
\be
{\cal H}(\mc{W},\La)={\cal H}({\cal W}/\Lambda)\,.
\ee
However, as the dependence on $\La$ must disappear upon integration over
superspace, we are lead uniquely to
\be
{\cal H}=\kappa\ln\frac{\cal W}{\Lambda}\,,
\label{log}
\ee
with some constant coefficient $\kappa$. Rescaling $\mc{W}$ then shifts
$\mc{H}$ by a constant which yields zero under the $d\zeta$ integral.

We conclude that the four-derivative part of the $\cN=4$ SYM effective
action on the Coulomb branch in $\cN=4$ $USp(4)$ harmonic superspace has
the following simple form
\be
\Ga_4=\kappa\int d\zeta dv\, \ln\frac{\cal  W}\Lambda\,.
\label{G4}
\ee
We will show that this action contains the `$F^4/X^4$' term (\ref{F4X4})
and the WZ term (\ref{WZso5}). This will allow us to fix the coefficient $\kappa$.

\subsection{The `$F^4/X^4$' term in the $\cN=4$ superspace action}

In order to identify the $F^4/X^4$ term (\ref{F4X4}) inside (\ref{G4}),
it is sufficient to consider $\mc{W}$ with \emph{constant} scalar fields
$\varphi$ and $X_a$. Then only the first line in (\ref{x15.2}) survives.
Substituting this simplified expression for $\mc{W}$ into the action
(\ref{S4}) and integrating over $\ta$'s, we find
\be
\Ga_{F^4}=\frac14\int d^4x dv\,{\cal H}^{(4)}(\varphi+i X_a v^5_a)
\Big(F_{mn}F^{nk}F_{kl}F^{lm}-\frac14(F_{pq}F^{pq})^2\Big)\,,
\label{S4_}
\ee
where $\mc{H}^{(n)}$ stands for the $n$'th derivative of $\mc{H}$ with respect to its argument.
To compute the harmonic integral, we expand ${\cal H}^{(4)}$ in the Taylor series,
\be
{\cal H}^{(4)}(\varphi+i X_a v^5_a)=\sum_{n=0}^\infty \frac1{n!} {\cal H}^{(4+n)}(\varphi)
(i X_a v^5_a)^n\,.
\label{ser}
\ee
Applying (\ref{206}) to each term in the series, we obtain
\be
\Ga_{F^4}=\frac14\int d^4x\,
\Big(F_{mn}F^{nk}F_{kl}F^{lm}-\frac14(F_{pq}F^{pq})^2\Big)
\sum_{n=0}^\infty \frac{3 (-X_a X_a)^n}{(2n+1)!(2n+3)}
{\cal H}^{(4+2n)}(\varphi)\,.
\label{S4.1}
\ee
For the function $\cal H$ given in (\ref{log}), we have
\be
{\cal H}^{(n)}(\varphi)=\kappa\frac{(-1)^{n-1}(n-1)!}{\varphi^n}\,.
\label{Hderiv}
\ee
Substituting this expression into (\ref{S4.1}) and summing the series, we find
\be
\Ga_{F^4}=-\frac32 \kappa\int d^4x\,
\frac{F_{mn}F^{nk}F_{kl}F^{lm}-\frac14(F_{pq}F^{pq})^2}
{(\varphi^2+X_a X_a)^2}\,.
\label{SF4}
\ee
This matches (\ref{F4X4}) provided we identify $\varphi=X_6$ and set
\be
\kappa=-\frac1{96\pi^2}\,.
\label{k}
\ee
Then the superfield action (\ref{G4}) contains the `$F^4/X^4$' term (\ref{F4X4}).

\subsection{The WZ term in the $\cN=4$ superspace action}

In order to identify the WZ term (\ref{WZso5}) inside (\ref{G4}), we keep the terms
in (\ref{x15.2}) with derivatives on the scalars, but ignore the terms
with $F_{m n}$. Substituting the resulting expression for $\mc{W}$ in
(\ref{G4}) and integrating over $\ta$'s, we then find
\bea
\Ga_4 &=&\int d^4xdv\,
{\cal H}^{(4)}(\varphi+i X_e v^5_e)
\partial^{\alpha\dot\alpha} X_a
\partial^{\beta\dot\beta}X_b
\partial_{\alpha\dot\beta}X_c
\partial_{\beta\dot\alpha}X_d\nn\\
&&\hspace{100pt}\times
(v^1_a-iv^2_a)(v^1_b+iv^2_b)
(v^3_c+iv^4_c)(v^3_d-iv^4_d)\nn\\
&-& \int d^4x dv\,{\cal H}^{(3)}(\varphi+i X_e v^5_e)
 \partial^{\alpha\dot\alpha}X_a \partial^{\beta\dot\beta}X_b
 \partial_{\alpha\dot\alpha}\partial_{\beta\dot\beta}(\varphi-i X_c v^5_c)
\nn\\&&\hspace{100pt}\times
 (v^1_a-iv^2_a)(v^1_b+iv^2_b)\nn\\
&-&\int d^4x dv\,{\cal H}^{(3)}(\varphi+i X_e v^5_e)
 \partial^{\alpha\dot\beta}X_a \partial^{\beta\dot\alpha}X_b
 \partial_{\alpha\dot\alpha}\partial_{\beta\dot\beta}(\varphi-i X_c v^5_c)
 \nn\\&&\hspace{100pt}\times
 (v^3_a+i v^4_a)(v^3_b-i v^4_b)\nn\\
&+&\frac12\int d^4x dv\, {\cal H}^{(2)}(\varphi+i X_e v^5_e)
\partial^{\alpha\dot\alpha}\partial^{\beta\dot\beta}
(\varphi-i X_a v^5_a)
\partial_{\alpha\dot\alpha}\partial_{\beta\dot\beta}
(\varphi-i X_b v^5_b)\,.\qquad
\label{xxx}
\eea
The Levi-Civita tensor, required for the WZ term, arises only from the cyclic contraction of the spinor indices on four $\p$'s. With $\p_{\al\dot\al}=\si_{\al\dot\al}^m\p_m$ and $\p^{\al\dot\al}=\tilde\si^{m\al\dot\al}\p_m$, the relevant trace formula is
\be
\tr\,\tilde\sigma^m\sigma^n\tilde\sigma^p
\sigma^q=-2i\varepsilon^{mnpq}
+2(\eta^{mn}\eta^{pq}+\eta^{np}\eta^{mq}-\eta^{mp}\eta^{nq})
\,.
\label{tr-sigma}
\ee
In addition, if two $\p$'s act on the same object, there is no contribution to the WZ term as $\eps^{m n p q}\p_m\p_n$ vanishes. Therefore, only the first integral in (\ref{xxx}) contributes, and we find
\be
\Ga_{WZ}=8i\varepsilon^{mnpq}\int d^4xdv\,{\cal H}^{(4)}(\varphi+i X_e v^5_e)
\partial_m X_a
\partial_n X_b
\partial_p X_c
\partial_q X_d
v^1_a v^2_b v^3_c v^4_d\,.
\label{SWZ}
\ee
Using again the series expansion (\ref{ser}) and computing the harmonic integral
for each term in the series with the help of (\ref{206}), we obtain
\be
\Ga_{WZ}=-
\varepsilon^{mnpq}\varepsilon^{abcde}\int d^4x\,
X_a
\partial_m X_b
\partial_n X_c
\partial_p X_d
\partial_q X_e
\sum_{n=0}^\infty\frac{(-X_fX_f)^n {\cal H}^{(2n+5)}(\varphi)}{
(2n+5)(2n+3)(2n+1)!}\,.
\label{SWZ.1}
\ee
Substituting (\ref{Hderiv}) into (\ref{SWZ.1}) and summing the series, we find
\be
\Ga_{WZ}=-\frac85 \kappa
\varepsilon^{mnpq}\varepsilon^{abcde}\int d^4x\,
\frac{g\left(\sqrt{\frac{X_f X_f}{\varphi^2}}\right)}{\varphi^5}
X_a
\partial_m X_b
\partial_n X_c
\partial_p X_d
\partial_q X_e
\,,
\label{SWZ.2}
\ee
where
\be
g(z)=\frac5{8 z^5}\left(
      3\arctan z-
          \frac{z
           \left( 3 +
          5z^2 \right) }
         {{\left( 1 + z^2
           \right) }^2}
 \right).
\ee
This matches (\ref{WZso5},\ref{gsum}) with $N=1$ perfectly, provided we once again identify
$\varphi=X_6$ and use the value for $\kappa$ given in (\ref{k}).

We have therefore established that the $\cN=4$ harmonic superspace action
(\ref{BLS}) contains both the `$F^4/X^4$' term (\ref{F4X4}) and the WZ term
(\ref{WZterm}) and therefore represents the four-derivative part in the
$\cN=4$ SYM effective action on the Coulomb branch.

\section{Effective action in $\cN=2$ harmonic superspace}
\label{sec-EAN2}

In this section, we will analyze the $\cN=2$ harmonic superspace form
of the four-derivative part in the $\cN=4$ SYM effective action on the
Coulomb branch. This form was found in \cite{BuIv} via $\cN=4$
supersymmetrization of the $\cN=2$ supersymmetric non-holomorphic
potential \cite{deWit:1996kc,DS}. We will establish that it does include
the WZ term (\ref{WZterm}), which this time will arise in its
$SO(4)\times SO(2)$ form (\ref{WZso4so2}).

\subsection{Scale-invariant and $\cN=4$ supersymmetric effective action}
The $\cN=4$ gauge multiplet consists of an $\cN=2$ gauge multiplet
and an $\cN=2$ hyper\-multiplet. Within the $\cN=2$ harmonic superspace
approach \cite{book}, these two multiplets are described
by off-shell unconstrained harmonic
superfields (see Appendix \ref{HSS2}). However, for the purpose of writing the
$\cN=4$ SYM effective action \emph{on-shell}, we will use \emph{constrained}
superfields. With all component fields satisfying their classical
equations of motion (\ref{oncshell}), the $\cN=2$ gauge superfield
strength $W$ (together with its conjugate $\bar W$) and the $\cN=2$
hypermultiplet superfield $q^+_a=(q^+,-\bar q^+)$
\footnote{In this section, we use indices $a,b,c=1,2$ for the
Pauli-Gursey $SU(2)$ group \cite{book}. These indices should not be confused
with the $SO(5)$ ones used in the previous section.}
have the following component expansions in the bosonic sector:
\bea
\label{W-comp}
{W}&=&\phi+2i\theta^-\sigma^m\bar\theta^+\partial_m \phi
+\frac1{\sqrt2}\theta^+_\alpha \theta^-_\beta
\sigma^{m\alpha}{}_{\dot\alpha}\sigma^{n\beta\dot\alpha}F_{mn}
\nn\\
\bar{W}&=&\bar\phi+2i\theta^+\sigma^m\bar\theta^-\partial_m \bar\phi
+\frac1{\sqrt2}\bar\theta^-_{\dot\beta}\bar\theta^+_{\dot\alpha}
\sigma^{m\dot\alpha}{}_{\alpha}\sigma^{n\alpha\dot\beta}F_{mn}
\eea
and
\bea
\label{q-comp}
q^{+} &=& f^{i}u^+_i+2i\theta^+\sigma^m\bar\theta^+\partial_m f^{i}u^-_i \nn\\[5pt]
\bar q^{+} &=& -\bar f^{i}u^+_i-2i\theta^+\sigma^m\bar\theta^+\partial_m \bar f^{i}u^-_i \ .
\eea
The six scalars are now described by complex fields $\phi$ and $f^i$ in the
${\bf 1}$ and ${\bf 2}$ of $SU(2)_R$, respectively. Under the remaining
$U(1)_R$ of the total $U(2)$ R-symmetry of the $\cN=2$ superalgebra,
$\ta_\al^\pm$ have charge $+1$, $\phi$ has charge $+2$, whereas $f^i$ and 
harmonics $u_i^\pm$ are neutral. (The $\pm$ on $\ta_\al^\pm$ and $u_i^\pm$ are charges
under a $U(1)$ subgroup of $SU(2)_R$.) On the other hand, the index $a$
on $q_a^{+}$ refers to a different $SU(2)$, the so-called Pauli-Gursey
$SU(2)_{P G}$. The $U(2)$ R-symmetry together with $SU(2)_{P G}$ gives rise
to the $SU(2)_{P G}\times SU(2)_R\times U(1)_R$ subgroup (\ref{msubs})
in the $SU(4)$ R-symmetry of the $\cN=4$ SYM.

The complete $\cN=4$ SYM effective action, $\Ga$, is a functional of
the superfields $W,\bar W$ and $q_a^{+}$.
In the four-derivative part,
these superfields must appear without derivatives,
\be
\Gamma_4=\int d^4x d^8\theta du \,{\cal L}_4({W},\bar{W},q^+,\bar q^+)\,,
\label{EA}
\ee
as the Grassmann part of the full $\cN=2$ superspace measure already provides the appropriate number of derivatives.
The part of $\mc{L}_4$ that depends only on $W$ and $\bar{W}$ is called the
`non-holomorphic potential' \cite{Henningson:1995eh}. Scale invariance fixes
the form of this potential uniquely, up to a coefficient
\cite{deWit:1996kc,DS}
\be
{\cal L}_4({W},\bar{W},q^+,\bar q^+)|_{q=0}=c\ln\frac{W}{\La}\ln\frac{\bar{W}}{\La}\,.
\label{non-hol}
\ee
Fixing the coefficient $c$ requires explicit one-loop calculations. For the case at hand, with $SU(2)$ gauge group spontaneously broken to $U(1)$, this yields \cite{non-hol1,non-hol2,non-hol3}
\bea
\label{c}
c=\frac{1}{(4\pi)^2} \ .
\eea

Buchbinder and Ivanov \cite{BuIv} showed that including the hypermultiplets in a way consistent with on-shell $\cN=4$ supersymmetry leads uniquely to
\be
{\cal L}_4({W},\bar{W},q^+,\bar q^+)=c\Big\{ \ln\frac{W}\Lambda
\ln\frac{\bar{W}}\Lambda+{H}(Z)\Big\}
\label{L}
\ee
with~\footnote{
The function (\ref{H}) has been rederived in \cite{BuIv-pert1,BuIv-pert2,BuIv-pert3,BuIv-pert4} through perturbative quantum calculations using powerful computational techniques of the off-shell $\cN=2$ harmonic superspace.
}
\be
{H}(Z)=
\frac{Z-1}{Z}\ln(1-Z)+{\rm Li}_2(Z)-1
=\sum_{n=1}^\infty \frac{Z^n}{n^2(n+1)}\,,
\label{H}
\ee
where Li$_2(Z)$ is the dilogarithm function and the superfield $Z$ is defined by
\be
Z=-\frac{q^{+a}q^-_a}{\bar{W}W}\,.
\label{X}
\ee
Here $q_a^{-}=D^{--}q_a^{+}$. This $Z$ is manifestly $SU(2)_{P G}\times SU(2)_R\times U(1)_R$ invariant.

\subsection{The `$F^4/X^4$' term in the $\cN=2$ superspace action}

In \cite{BuIv}, it has been verified that the superfield expression for $\Ga_4$ contains the `$F^4/X^4$' term (\ref{F4X4}). Indeed, using (\ref{W-comp}) and (\ref{q-comp}) with constant scalars, and performing the superspace integration in (\ref{EA}) with $\mc{L}_4$ given in (\ref{L}), we find
\bea
\Ga_{F^4} &=& \frac{c}{4}\int d^4x
\frac{F_{m n}F^{n k}F_{k l}F^{l m}-\qter(F_{p q}F^{p q})}{\phi^2\bar\phi^2}\sum_{n=0}^\infty
(n+1)\left(\frac{-f^i\bar f_i}{\phi\bar\phi}\right)^n \nn\\
&=& \frac{c}{4}\int d^4x
\frac{F_{m n}F^{n k}F_{k l}F^{l m}-\qter(F_{p q}F^{p q})}{(\phi\bar\phi+f^i\bar f_i)^2} \ .
\eea
With $c$ given in (\ref{c}), and $f^i$ and $\phi$ related to the six real scalars $X_A$ via
\bea
\label{fphiX}
f^1=X_1+i X_2, \quad f^2=X_3+i X_4, \quad \phi=X_6+i X_5; \quad
\bar{f}_i=\ov{(f^i)}, \quad \bar{\phi}=\ov{(\phi)} \ ,
\eea
this yields the coefficient of the `$F^4/X^4$' term quoted in (\ref{F4X4}).

\subsection{The WZ term in the $\cN=2$ superspace action}

For a general $\mc{L}_4$, the action (\ref{EA}) contains a pure scalar WZ-like
term \cite{AABE}. As we will show next, the particular $\mc{L}_4$ given in
(\ref{L}) leads to the $SO(6)$-invariant WZ term (\ref{WZterm}).

The non-holomorphic potential (\ref{non-hol}) does not contribute to
the WZ term, as it depends only on two of the six scalars. Therefore,
the WZ term is determined by the function $H(Z)$ in (\ref L). 
We evaluate the relevant integral, $\int d^4\ta H(Z)$, by projection with the superspace covariant derivatives, see Appendix \ref{B.2}, and find the following result for the part involving the Levi-Civita tensor,
\bea
\Ga_{WZ} &=& 2i c\varepsilon^{mnpq}\int d^4x du\bigg[
\frac{\partial^4{H}(z)}{\partial f^+_a \partial f^+_b \partial f^-_c \partial f^-_d}
\partial_m f^-_d \partial_n f^+_c
\partial_p f^+_b
\partial_q f^-_a
\nn\\
&&\hspace{-50pt}
+\frac{\partial^4{H}(z)}{\partial\phi\partial f^+_a\partial f^+_b\partial f^-_c}
\partial_m \phi
\partial_n f^+_c \partial_p f^+_b
\partial_q f^-_a
+\frac{\partial^4{H}(z)}{\partial\phi\partial f^+_a\partial f^-_c \partial f^-_d}
\partial_m f^-_d \partial_n f^+_c
\partial_p \phi
\partial_q f^-_a\bigg]
\,, \quad
\label{l135}
\eea
where $f_a^{\pm}=f_a^i u_i^{\pm}$ are the lowest components of $q_a^{\pm}$, and
\be
z=\frac{f^+_a\varepsilon^{ab} f^{-}_b}{\phi\bar\phi}
=-\frac{ f^i \bar f_i}{\phi\bar\phi}=-\frac{X_1^2+X_2^2+X_3^2+X_4^2}{X_5^2+X_6^2}
\label{z}
\ee
is the lowest component of the superfield $Z$ in (\ref{X}). The $\Ga_{WZ}$ in (\ref{l135}) is the sum of three terms, $\Ga_{WZ}=T_1+T_2+T_3$, of which the first one is imaginary and the other two are complex. The imaginary part of $T_2+T_3$ can be written as follows
\bea
\label{l137}
i c\varepsilon^{mnpq}\int d^4x du\, \partial_n f^+_c
\partial_q f^-_a(\partial_p f^+_b \frac\partial{\partial f^+_b}-\partial_p f^-_b
 \frac\partial{\partial f^-_b})
 (\partial_m\phi \frac\partial{\partial\phi}+\partial_m\bar\phi\frac\partial{\partial
 \bar \phi})
 \frac{\partial^2{H}(z)}{\partial f^+_a\partial f^-_c}\,.
\eea
Integrating here by parts with $\p_m$, we obtain precisely $-T_1$, i.e. the first term in (\ref{l135}) with opposite sign. Therefore, $\Ga_{WZ}$ is equal to the real part of $T_2+T_3$,~\footnote{
Contrary to eq.~(6.14) in \cite{AABE}, we find it impossible to rewrite $\Ga_{WZ}$ so that $\phi$ and $\bar\phi$ appear without derivatives. Their eq.~(6.12), however, agrees with our eq.~(\ref{l135}). On the other hand, eq.~(4.13) in \cite{BuIv-pert4} is incomplete, as it includes only the first (imaginary) term in our eq.~(\ref{l135}). Of course, these small corrections do not affect the key results of \cite{AABE} and \cite{BuIv-pert4}.
}
\bea
\label{A14}
i c\varepsilon^{mnpq}\int d^4x du\, \partial_n f^+_c
\partial_q f^-_a
 (\partial_p f^+_b \frac\partial{\partial f^+_b}-\partial_p f^-_b
  \frac\partial{\partial f^-_b})
 (\partial_m\phi\frac\partial{\partial \phi}
  -\partial_m\bar\phi\frac\partial{\partial\bar{\phi}})
  \frac{\partial^2{H}(z)}{\partial f^+_a \partial f^-_c}\,.
\eea
The partial derivatives of the function $H$ with respect to
$f^\pm_a$, $\phi$ and $\bar\phi$ reduce to ordinary derivatives with respect to $z$. As $z$, given in (\ref{z}), is independent of the harmonics, the identities (\ref{B9}) are sufficient to evaluate all the harmonic integrals. This way we find
\bea
\Ga_{WZ} &=& i c\varepsilon^{mnpq}\int d^4x\, \left(\frac{\p_m\phi}{\phi}-\frac{\p_m\bar\phi}{\bar\phi}\right)
\bigg\{
\partial_q f^i_a \partial_n f_i^c \partial_p f^j_c f_j^a
\frac{2{H}^{(2)}+z{H}^{(3)}}{(\phi\bar\phi)^2}\nn\\
&&\hspace{-35pt}
-\left(\frac1{12}\partial_n f^i_c f^c_k \partial_q f^k_a f^a_j
 \partial_p f^j_b f^b_i
 +\frac18 f^{ak}f_{ak}\partial_n f^i_c \partial_q f_j^c
  \partial_p f^j_b f_i^b\right)\frac{3{H}^{(3)}+z{H}^{(4)}}{(\phi\bar\phi)^3}
 \bigg\},
 \label{l142}
\eea
where ${H}^{(n)}={d^n {H}(z)}/{d z^n}$. With $f^{i}_a=(f^i,\bar f^i)$ and $f_i^a=(-\bar f_i,f_i)$, we then obtain
\be
\Ga_{WZ}=i c\varepsilon^{mnpq}\int d^4x\,
\Big[6{H}^{(2)}+6z{H}^{(3)}+z^2{H}^{(4)}\Big]
\frac{\partial_n f^i\partial_p \bar f_i(
  \partial_q f^j \bar f_j -\partial_q \bar f_j f^j)}{(\phi\bar\phi)^2}
\partial_m\ln\frac{\phi}{\bar\phi}\,.
\label{l144}
\ee
Finally, using (\ref{fphiX}) together with the polar decomposition for $\phi$,
\bea
\phi=X_6+i X_5=X e^{i\al} \ ,
\eea
we find that
\be
\Ga_{WZ}=-\frac{4 c}3\varepsilon^{mnpq}\varepsilon^{a'b'c'd'}
\int d^4x\, \Big[6{H}^{(2)}+6z{H}^{(3)}+z^2{H}^{(4)}\Big]
\frac{X_{a'} \partial_n X_{b'} \partial_p X_{c'} \partial_q X_{d'}}{X^4}\partial_m\al
\,,
\label{l150}
\ee
where $a',b'=1,2,3,4$ are $SO(4)$ indices
and $\varepsilon^{1234}=1$.
For any $H(z)$, this WZ-like term has manifest $SO(4)\times SO(2)$ invariance. But only for particular $H(z)$ this invariance gets extended to $SO(6)$. The $H(z)$ given in (\ref{H}) is one such function. It satisfies~\footnote{
The function (\ref{H}) is a particular solution to the fourth order differential equation (\ref{difur}). The general solution is a sum of this particular solution and
\be
\frac{c_1}{z}+c_2+c_3\ln z+ c_4 z \ , \nn
\ee
where the four $c$'s are arbitrary constants. Therefore, requiring that (\ref{EA}) with (\ref{L}) yields the WZ term (\ref{WZfromN2}) does not fix $H(z)$ uniquely. However, as shown in \cite{BuIv}, the requirement of $\cN=4$ supersymmetry selects the $H(z)$ in (\ref{H}) as the unique possibility.
}
\be
6{H}^{(2)}(z)+6z{H}^{(3)}(z)+z^2{H}^{(4)}(z)=\frac1{(z-1)^2}\ ,
\label{difur}
\ee
so that (\ref{l150}) becomes
\be
\label{WZfromN2}
\Ga_{WZ}=\frac43 c\,\varepsilon^{mnpq}\varepsilon^{a'b'c'd'}
\int d^4x\,
\frac{X_{a'} \partial_m X_{b'} \partial_n X_{c'} \partial_p X_{d'}
}{(X_{e'} X_{e'}+X^2)^2}\partial_q \alpha\ .
\ee
With $c$ given in (\ref{c}), this matches (\ref{WZso4so2}) perfectly.

Therefore, we have explicitly verified that the action of Buchbinder
and Ivanov \cite{BuIv}, providing the $\cN=2$ harmonic superspace
expression for $\Ga_4$, contains both the `$F^4/X^4$' term
(\ref{F4X4}) and the WZ term (\ref{WZterm}).

\section{Summary and discussion}
\label{sec-summary}

In this paper, we discussed the Wess-Zumino term \cite{TZ,Intriligator:2000eq} in the low-energy effective action for $\cN=4$ SYM on the Coulomb branch. The WZ term has well-known five-dimensional form (\ref{WZterm}) with manifest $SO(6)$ R-symmetry. We found, however, that it is also important to know its four-dimensional forms, even though the full $SO(6)$ symmetry cannot be manifest in such a formulation. We argued that the subgroups of $SO(6)$ that \emph{can} be made manifest determine natural superspaces for the description of the effective action.

As the WZ term reflects the anomaly in the R-symmetry currents, the determining factor is whether the subgroups are anomalous or not. We found that three maximal subgroups, $SO(5)$, $SO(4)\times SO(2)$, and $SO(3)\times SO(3)$, are non-anomalous, and we explicitly constructed three four-dimensional forms of the WZ term with these symmetries being manifest. The $SO(5)$ form has been discussed before \cite{Braaten:1985is,Claus:1998mw}, whereas the two other forms are new. (We also demonstrated that in a general $d$-dimensional $SO(d+2)$-invariant WZ term \cite{Braaten:1985is} the $SO(n)\times SO(d+2-n)$ subgroup for any $n$ can be made manifest.) The fourth maximal subgroup of $SO(6)\simeq SU(4)$ R-symmetry, $SU(3)\times U(1)$, is however anomalous, which implies that it is \emph{not} possible to keep it manifest in a four-dimensional form of the WZ term.

We showed that the $SO(5)$ and $SO(4)\times SO(2)$ R-symmetry subgroups point naturally to $\cN=4$ $USp(4)$ \cite{IKVO,BLS} and $\cN=2$ $SU(2)$ \cite{book} harmonic superspaces, respectively. Starting with the known expressions \cite{BLS,BuIv} for the $\cN=4$ SYM effective action in these superspaces, we identified WZ-like terms that they contain and found that these match perfectly the $SO(5)$ and $SO(4)\times SO(2)$ forms of the $SO(6)$-invariant WZ term. In the $\cN=2$ case \cite{BuIv}, our results correct and complete similar investigations in \cite{AABE,BuIv-pert4}. The WZ term in the $\cN=4$ formulation \cite{BLS} has not been previously discussed.

Our results also explicitly confirm that the $\cN=4$ supersymmetrization of either the `$F^4/X^4$' term (\ref{F4X4}) or the WZ term (\ref{WZterm}) leads to the same action, which is the four-derivative part in the $\cN=4$ SYM effective action \cite{DS}.

The forms of the $\cN=4$ SYM effective action in the $\cN=2$ \cite{BuIv} and $\cN=4$ \cite{BLS} superspaces were found as unique superfield expressions obeying the requirements of scale invariance and full $\cN=4$ supersymmetry. These symmetries leave only the overall coefficient undetermined. Fixing this coefficient requires explicit one-loop calculations. However, as the coefficient in front of the WZ term must be quantized \cite{Witten:1983tw}, the ambiguity is reduced to choosing an integer $N$ in (\ref{WZso6}).
In the simplest case that we considered, with $SU(2)$ gauge group spontaneously broken to $U(1)$, this coefficient takes its minimal allowed value, $N=1$ \cite{TZ}. In the general case of a gauge group $G_\text{g}$ broken to its subgroup $H_\text{g}$, one finds $N=(|G_\text{g}|-|H_\text{g}|)/2$ \cite{Intriligator:2000eq}. 

Our results also shed more light on the problem of describing the $\cN=4$ SYM effective action in $\cN=3$ harmonic superspace and in the conventional $\cN=1$ superspace.

The $\cN=3$ harmonic superspace \cite{Galperin:1985uw,Galperin:1984bu} is based on the $SU(3)$ R-symmetry group. However, as we argued, it is not possible to write the WZ term in a four-dimensional form with $SU(3)$ R-symmetry being manifest. This makes the formulation of the effective action in the $\cN=3$ superspace non-trivial. The $SU(3)$ R-symmetry must be explicitly broken in the superspace Lagrangian, but in such a way that it is restored upon superspace integration. This would be similar to the way the constant $\La$ apparently breaks scale invariance in (\ref{BLS}) and (\ref{DS+BUIV}). Alternatively, one can have manifest $SU(3)$ R-symmetry at the price of non-manifest locality. Such a form of the $\cN=4$ SYM effective action has been proposed in \cite{Buchbinder:2004rj}. Similarly, one could try to maintain manifest $SU(4)$ R-symmetry by either sacrificing manifest locality in four dimensions, or by $\cN=4$ supersymmetrizing the WZ term (\ref{WZterm}) directly in five dimensions \cite{Rohm}.

In the $\cN=1$ superspace formulation of the classical $\cN=4$ SYM action, the $SU(3)$ R-symmetry rotates three chiral superfields \cite{GGRS}, whereas in the $\cN=3$ superspace it rotates Grassmann coordinates. Nonetheless, we conclude that in the $\cN=1$ superspace formulation of the effective action \cite{DS}, the $SU(3)$ R-symmetry cannot be manifest. This makes the problem of constructing the $\cN=1$ form of the effective action $\Ga_4$ particularly interesting.

Finally, the form of the WZ term with manifest $SO(3)\times SO(3)$ subgroup of the $SO(6)$ R-symmetry deserves further study. It would be interesting to see to which superspace formulation of the $\cN=4$ SYM effective action does it correspond.

\vspace{30pt}
\noindent
{\bf Acknowledgments}\\[3mm]
D.B.\ thanks J.~Bagger, S.~Ketov, S.~Kuzenko, P.~Ramond, W.~Siegel
and A.~Tseytlin for helpful discussions and correspondence. The
research of D.B. was supported by the Department of Energy Grant
No.\ DE-FG02-97ER41029. I.S.\ is grateful to I.L.~Buchbinder and
D.~Sorokin for stimulating discussions. The work of I.S.\ was
supported by the Marie Curie research fellowship No 236231,
``QuantumSupersymmetry''; by RFBR grants Nr 09-02-00078 and
11-02-90445; by grant for LRSS, project No 3558.2010.2 as well as
by the RF Federal programm ``Kadry'' contract Nr P691.

\appendix

\section{$\cN=4$ $USp(4)$ harmonic superspace}
\label{HSS4}

$\cN=4$ $USp(4)$ harmonic superspace has been developed in \cite{IKVO,BLS}.
Here we will give a summary of its basic features relevant for the
discussion in Section~\ref{sec-EAN4}.

\subsection{$USp(4)$ harmonics and covariant derivatives}

Harmonics $u^\ui{}_i$ on a coset $G/H$ correspond to a $G$-matrix with $i$ running over the fundamental representation of $G$ and $\ui$ running over the reducible representation of $H$ that defines its embedding in $G$ \cite{IKVO}. For $G=USp(4)$ and $H=U(1)\times U(1)$ embedded as
\bea
{\bf 4}=(+1,0)+(-1,0)+(0,+1)+(0,-1) \,,
\eea
this gives $USp(4)/[U(1)\times U(1)]$ harmonics $u^{\ui}{}_i$, with $i=1,2,3,4$ and
\be
\ui=(+,0),(-,0),(0,+),(0,-) \,,
\label{I}
\ee
which form an $USp(4)$ matrix $u$,
\be
u u^\dag=I_4\,,\qquad u\Omega u^{\rm T}=\Omega\,.
\label{USp}
\ee
Here $I_4$ is the unit matrix and $\Omega$ is a constant antisymmetric
matrix, $\Omega^{\rm T}=-\Omega$. Being the $USp(4)$ invariant tensor,
$\Om$ is used to raise and lower $USp(4)$ indices, e.g.,
\be
u^{\ui i}=\Omega^{ij}u^\ui{}_j\,,\quad
u^\ui{}_i=\Omega_{ij}u^{\ui j}\,;\quad
\Omega^{ij}\Omega_{jk}=\delta^i_k\,.
\ee
Constraints (\ref{USp}) yield orthogonality conditions,
\bea
&&u^{(+,0)i}u^{(-,0)}_i=u^{(0,+)i}u^{(0,-)}_i=1\,,\nn\\
&&u^{(+,0)}_i u^{(0,+)i}=u^{(+,0)}_i u^{(0,-)i}=
u^{(0,+)}_i u^{(-,0)i}=u^{(-,0)}_i u^{(0,-)i}=0\,,
\eea
and completeness relations,
\be
u^{(+,0)}_i u^{(-,0)}_j-u^{(+,0)}_j u^{(-,0)}_i+
u^{(0,+)}_i u^{(0,-)}_j-u^{(0,+)}_j u^{(0,-)}_i=\Omega_{ij}\,.
\ee
Grassmann coordinates $\theta_{i\alpha}$, $\bar\theta^i_{\dot\alpha}$ and the corresponding covariant spinor derivatives $D^i_\alpha$, $\bar D_{i\dot\alpha}$ of the conventional $\cN=4$ superspace are projected with the harmonics,
\be
\theta^\ui_\alpha=-u^{\ui i}\theta_{i\alpha}\,,\quad
\bar\theta^\ui_{\dot\alpha}=u^\ui_i\bar\theta^i_{\dot\alpha}\,,\quad
D^\ui_\alpha=u^\ui_i D^i_\alpha\,,\quad
\bar D^\ui_{\dot\alpha}=-u^{\ui i}\bar D_{i\dot\alpha}\,.
\ee

Complex conjugation operates as follows
\bea
\ov{(u_i^{(\pm,0)})}=\mp u^{(\mp,0)i}, \quad
\ov{(u_i^{(0,\pm)})}=\mp u^{(0,\mp)i}, \quad
\ov{(\Om_{i j})}=-\Om^{i j} \ .
\eea
Another useful `tilde-conjugation' is defined by
\bea
\label{tilde}
&&\widetilde{u^{(\pm,0)}_i}=u^{(0,\pm)i}\,,\quad
\widetilde{u^{(0,\pm)}_i}=u^{(\pm,0)i}\,,\quad
\widetilde{u^{(\pm,0)i}}=-u^{(0,\pm)}_i\,,\quad
\widetilde{u^{(0,\pm)i}}=-u^{(\pm,0)}_i\,,
\nn\\
&&\widetilde{\theta^{(\pm,0)}_\alpha}=\bar\theta^{(0,\pm)}_{\dot\alpha}\,,
\quad
\widetilde{\theta^{(0,\pm)}_\alpha}=\bar\theta^{(\pm,0)}_{\dot\alpha}\,,
\quad
\widetilde{\bar\theta^{(0,\pm)}_{\dot\alpha}}=-\theta^{(\pm,0)}_\alpha\,,
\quad
\widetilde{\bar\theta^{(\pm,0)}_{\dot\alpha}}=-\theta^{(0,\pm)}_\alpha\,,
\nn\\
&&\widetilde{D^{(\pm,0)}_\alpha}=-\bar
D^{(0,\pm)}_{\dot\alpha}\,,\quad
\widetilde{D^{(0,\pm)}_\alpha}=-\bar
D^{(\pm,0)}_{\dot\alpha}\,,\quad
\widetilde{\bar
D^{(\pm,0)}_{\dot\alpha}}=D^{(0,\pm)}_\alpha\,,\quad
\widetilde{\bar D^{(0,\pm)}_{\dot\alpha}}=D^{(\pm,0)}_\alpha\,. \qquad
\eea
Besides spinorial $D$'s, there are also bosonic $USp(4)$-covariant harmonic derivatives
\be
\begin{array}[b]{lll}
D^{(\pm\pm,0)}=u^{(\pm,0)}_i\dfrac{\partial}{\partial
u^{(\mp,0)}_i}\,,&&
D^{(0,\pm\pm)}=u^{(0,\pm)}_i\dfrac{\partial}{\partial
u^{(0,\mp)}_i}\,,\\
D^{(\pm,\pm)}=u^{(\pm,0)}_i\dfrac\partial{\partial u^{(0,\mp)}_i}
+u^{(0,\pm)}_i\dfrac\partial{\partial u^{(\mp,0)}_i}\,,&&
D^{(\pm,\mp)}=u^{(\pm)}_i\dfrac\partial{\partial u^{(0,\pm)}_i}
-u^{(0,\mp)}_i\dfrac\partial{\partial u^{(\mp,0)}_i}\,,\\
S_1=u^{(+,0)}_i\dfrac\partial{\partial u^{(+,0)}_i}
 -u^{(-,0)}_i\dfrac\partial{\partial u^{(-,0)}_i}\,,&&
S_2=u^{(0,+)}_i\dfrac\partial{\partial u^{(0,+)}_i}
 -u^{(0,-)}_i\dfrac\partial{\partial u^{(0,-)}_i}\,.
\end{array}
\label{D}
\ee
The operators $S_1$ and $S_2$ measure the $U(1)\times U(1)$ charges of other operators,
\be
[S_1,D^{(s_1,s_2)}]=s_1 D^{(s_1,s_2)}\,,\qquad
[S_2,D^{(s_1,s_2)}]=s_2 D^{(s_1,s_2)}\,,\qquad
[S_1,S_2]=0 \ .
\label{u1u1}
\ee
The complete algebra of the harmonic derivatives is given in
eq.~(A.3) of \cite{BLS}. It is isomorphic to the Lie algebra of $USp(4)$. The two triplets of operators in
\bea
\label{su2su2}
[D^{(++,0)},D^{(--,0)}]=S_1\,, \quad
[D^{(0,++)},D^{(0,--)}]=S_2\,,
\eea
define an $SU(2)\times SU(2)$ subgroup of $USp(4)$.

\subsection{Analytic subspace}

$\cN=4$ $USp(4)$ harmonic superspace $\{x^m,\theta^\ui_{\alpha},\bar\theta^\ui_{\dot\alpha},u^{\ui}{}_i\}$ contains several analytic subspaces with 8 (out of the total 16) real Grassmann coordinates \cite{BLS}. One such subspace is parametrized by
\be
\{\zeta,u\}=\{(x^m_{[A]},\theta^{(+,0)}_\alpha,\theta^{(-,0)}_\alpha,
\bar\theta^{(0,+)}_{\dot\alpha},\bar\theta^{(0,-)}_{\dot\alpha}),u^{\ui}{}_i\}\,,
\label{ancor}
\ee
where
\be
x^m_{[A]}=x^m-i\theta^{(0,-)}\sigma^m\bar\theta^{(0,+)}
+i\theta^{(0,+)}\sigma^m \bar\theta^{(0,-)}
-i\theta^{(+,0)}\sigma^m\bar\theta^{(-,0)}
+i\theta^{(-,0)}\sigma^m \bar\theta^{(+,0)}\,.
\label{N4anal}
\ee
In this analytic subspace, the following Grassmann derivatives become short,
\be
D^{(0,\pm)}_{[A]\alpha}
=\pm\frac\partial{\partial\theta^{(0,\mp)\alpha}}\,,\qquad
\bar D^{(\pm,0)}_{[A]\dot\alpha}=\pm\frac\partial{
\partial\bar\theta^{(\mp,0)\dot\alpha}}\ .
\label{Dshort}
\ee
The analytic measure used for integrating over the analytic subspace (\ref{ancor}) is
\bea
d\zeta du=d^4 x_{[A]} d^8\ta_{[A]} du \ .
\label{meas}
\eea
Integration over Grassmann variables is defined by
\bea
\int d^8\ta_{[A]} (\theta^{(+,0)})^2(\theta^{(-,0)})^2
(\bar\theta^{(0,+)})^2 (\bar\theta^{(0,-)})^2=1\,,
\label{N4measure}
\eea
whereas the harmonic integral is defined to select the $USp(4)$ singlet
\bea
\int du\,1=1\,,\qquad \int du\, (\mbox{non-singlet $USp(4)$ irreducible representation})=0\,.
\label{N4hint}
\eea

\subsection{Gauge superfield strength}

In the conventional $\cN=4$ superspace $\{x^m,\theta_{i\alpha},\bar\theta^i_{\dot\alpha}\}$, the gauge superfield strength $W^{i j}$ is constrained by
\bea
\label{Wij}
&& W^{i j}=-W^{j i}, \quad \ov{W^{i j}}=\half\eps_{i j k l}W^{k l} \nn\\
&& D_\al^i W^{j k}+D_\al^j W^{i k}=0, \quad
\bar D_{i\dot\al}W^{j k}=\frac{1}{3}(\da_i^j\bar D_{l\al}W^{l k}-\da_i^k\bar D_{l\dot\al}W^{l j}) \ .
\eea
Among its harmonic projections, $W^{\ui\uj}=u^\ui{}_i u^\uj{}_j W^{ij}$, we select the following one~\footnote{
$\mc{W}$ is neutral with respect to the $U(1)\times U(1)$ subgroup of $USp(4)$: $S_1\mc{W}=S_2\mc{W}=0$.
}
\be
{\cal W}=u^{(0,+)}_i u^{(0,-)}_j W^{ij}\ .
\label{proj}
\ee
This superfield alone is sufficient to describe the $\cN=4$ gauge multiplet \cite{BLS}. The constraints (\ref{Wij})
imply the following restrictions on $\cal W$,
\bea
&&\widetilde{\cal W}={\cal W}\,,
\label{e119.2} \\
&&D^{(0,+)}_\alpha{\cal W}=D^{(0,-)}_\alpha{\cal W}
=\bar D^{(+,0)}_{\dot\alpha}{\cal W}
=\bar D^{(-,0)}_{\dot\alpha}{\cal W}=0\,,
\label{l9}\\
&&D^{(++,0)}{\cal W}=D^{(--,0)}{\cal W}=D^{(0,++)}{\cal W}=D^{(0,--)}{\cal W}=0\,,\label{l15.1}\\
&&(D^{(+,+)})^2{\cal W}=0\ .
\label{l15}
\eea
According to (\ref{e119.2}), $\mc{W}$ is real with respect to tilde-conjugation (\ref{tilde}). Thanks to (\ref{Dshort}), constraints (\ref{l9}) are trivially solved in the analytic subspace (\ref{ancor}):
\be
{\cal W}={\cal W}(x^m_{[A]},\theta^{(\pm,0)}_\alpha,\bar\theta^{(0,\pm)}_{\dot\alpha},u^{\ui}{}_i)\,.
\ee
In this analytic subspace, harmonic derivatives in (\ref{l15.1}) and (\ref{l15}) take the following form (omitting terms that act trivially on $\mc{W}$)
\bea
\label{Dharm1}
D^{(\pm\pm,0)}_{[A]}&=&D^{(\pm\pm,0)}+\theta^{(\pm,0)}\frac\partial{\partial
 \theta^{(\mp,0)}}\,,\qquad
D^{(0,\pm\pm)}_{[A]}=D^{(0,\pm\pm)}+\theta^{(0,\pm)}
 \frac\partial{\partial\theta^{(0,\mp)}}\,,
\\
\label{Dharm2}
D^{(+,+)}_{[A]} &=& D^{(+,+)}
+\theta^{(0,+)}\frac\partial{\partial\theta^{(-,0)}}
+\bar\theta^{(+,0)}\frac\partial{\partial\bar\theta^{(0,-)}} \nn\\[3pt]
&&\hspace{30pt} -2i(\theta^{(+,0)}\sigma^m\bar\theta^{(0,+)}
 -\theta^{(0,+)}\sigma^m\bar\theta^{(+,0)})\partial_{[A]m} \ ,
\eea
where $D^{(\pm\pm,0)}$, $D^{(0,\pm\pm)}$ and $D^{(+,+)}$ are as defined in (\ref{D}).
As the harmonic derivatives in (\ref{Dharm1}) do not involve space-time derivatives, constraints (\ref{l15.1}) are also kinematical. With (\ref{su2su2}), they simply express the fact that $\cal W$ depends only on $USp(4)/[SU(2)\times SU(2)]$ harmonics. The only dynamical constraint, which puts $\mc{W}$ on shell, is (\ref{l15}).

The superfield $\mc{W}$ describes the $\cN=4$ gauge multiplet with six scalars,
four spinors and one gauge field. The latter enters $\mc{W}$ as the gauge-invariant
field strength $F_{m n}$. The six scalars split into {\bf 1} and {\bf 5} of $USp(4)$
described, respectively, by $\varphi$ and $f^{i j}$ satisfying
\bea
\ov\varphi=\varphi; \quad f^{i j}=-f^{j i}, \quad f^{i j}\Om_{i j}=0, \quad \ov{(f^{i j})}=\bar f_{i j}=-f_{i j} \ .
\eea
Omitting the fermions (gauginos), the component expansion of $\mc{W}$ reads
\bea
\cal W&=&\varphi+f^{ij}(u^{(+,0)}_{[i}u^{(-,0)}_{j]}-u^{(0,+)}_{[i}u^{(0,-)}_{j]})
\nn\\&& +\frac1{\sqrt2} (\theta^{(+,0)}_\alpha\theta^{(-,0)}_\beta
\sigma^{m\alpha}{}_{\dot\alpha}\sigma^{n\beta\dot\alpha}
-\bar\theta^{(0,+)}_{\dot\alpha}\bar\theta^{(0,-)}_{\dot\beta}
\sigma^{m\dot\alpha}{}_\alpha \sigma^{n\alpha\dot\beta})F_{mn}
\nn\\&& -4i\theta^{(+,0)}_\alpha\bar\theta^{(0,+)}_{\dot\alpha}
\partial^{\alpha\dot\alpha}f^{ij}u^{(-,0)}_{[i}u^{(0,-)}_{j]}
-4i\theta^{(-,0)}_\alpha\bar\theta^{(0,-)}_{\dot\alpha}
\partial^{\alpha\dot\alpha}f^{ij}u^{(+,0)}_{[i}u^{(0,+)}_{j]}
\nn\\&& +4i\theta^{(+,0)}_\alpha\bar\theta^{(0,-)}_{\dot\alpha}
\partial^{\alpha\dot\alpha}f^{ij}u^{(-,0)}_{[i}u^{(0,+)}_{j]}
+4i\theta^{(-,0)}_\alpha\bar\theta^{(0,+)}_{\dot\alpha}
\partial^{\alpha\dot\alpha}f^{ij}u^{(+,0)}_{[i}u^{(0,-)}_{j]}
\nn\\&& +4\theta^{(+,0)}_\alpha\theta^{(-,0)}_\beta
\bar\theta^{(0,+)}_{\dot\alpha}\bar\theta^{(0,-)}_{\dot\beta}
\partial^{\alpha\dot\alpha}\partial^{\beta\dot\beta}
[\varphi-f^{ij}(u^{(+,0)}_{[i}u^{(-,0)}_{j]}-u^{(0,+)}_{[i}u^{(0,-)}_{j]})] \ .
\label{x15_}
\eea
The missing gaugino-dependent terms are given in eq.~(5.21) of \cite{BLS}.

\subsection{$SO(5)$ harmonics}

The ${\bf 5}$ of $USp(4)\simeq SO(5)$ is given by the antisymmetric
$\Om$-traceless part of ${\bf 4}\times{\bf 4}$. The corresponding
Clebsch-Gordan coefficients are gamma matrices $\gamma_a^{i j}$, with
$a=1,2,3,4,5$ of $SO(5)$ and $i=1,2,3,4$ of $USp(4)$, such that
\bea
&& \gamma_a^{ij}=-\gamma_a^{j\,i}, \quad \Omega_{i j}\gamma_a^{i j}=0, \quad
\ga_{a i j}\ga_b^{j k}+\ga_{b i j}\ga_a^{j k}=2\da_{ab}\da_i^k, \quad
\ov{(\ga_a^{i j})}=-\ga_{a i j} \,, \nn\\[3pt]
&& \gamma_a^{ij}\gamma_{b\,ij}=-4\delta_{ab}\,,\quad
\gamma_{a i j}\gamma_a^{k l}=-2(\delta_i^k\delta_j^l-\delta_i^l\delta_j^k)
-\Omega_{ij}\Omega^{kl}\,.
\eea
Using the bilinear combinations of $USp(4)/[U(1)\times U(1)]$ harmonics appearing in (\ref{x15_}), we define
\bea
&&v^{(-,-)}_a = \gamma_a^{ij}u^{(-,0)}_{[i}u^{(0,-)}_{j]}, \quad
v^{(+,+)}_a = \gamma_a^{ij}u^{(+,0)}_{[i}u^{(0,+)}_{j]}
\,,\nn\\
&&v^{(-,+)}_a = \gamma_a^{ij}u^{(-,0)}_{[i}u^{(0,+)}_{j]}, \quad
v^{(+,-)}_a = \gamma_a^{ij}u^{(+,0)}_{[i}u^{(0,-)}_{j]}
\,, \nn\\
&&v^{(0,0)}_a = \gamma^{ij}_a(u_{[i}^{(+,0)}u^{(-,0)}_{j]}-u_{[i}^{(0,+)}u^{(0,-)}_{j]})
\ .
\label{so5u1u1h}
\eea
These objects have definite $U(1)\times U(1)$ charges \cite{Z1}, but they do not form an $SO(5)$ matrix as their non-zero products are
\bea
v_a^{(-,-)}v_a^{(+,+)}=-2, \quad
v_a^{(-,+)}v_a^{(+,-)}=+2, \quad
v_a^{(0,0)}v_a^{(0,0)}=-4 \ .
\eea
We therefore define $SO(5)$ harmonics $v_b^a$ by
\bea
&& v_a^1=\frac{1}{2}(v_a^{(-,-)}-v_a^{(+,+)}), \quad
v_a^2=\frac{i}{2}(v_a^{(-,-)}+v_a^{(+,+)}) \nn\\
&& v_a^3=\frac{i}{2}(v_a^{(-,+)}-v_a^{(+,-)}), \quad
v_a^4=\frac{1}{2}(v_a^{(-,+)}+v_a^{(+,-)}), \quad
v_a^5=-\frac{i}{2}v_a^{(0,0)} \ .
\label{cor-harm}
\eea
These are real, $\ov{(v_a^b)}=v_a^b$, and obey
\be
v^a_c v^b_c=\delta^{ab}\,,\qquad
\varepsilon^{abcde}v^1_a v^2_b v^3_c v^4_d v^5_e=1\ .
\ee
The integration over $SO(5)$ harmonic variables is defined by
\be
\int dv\,1=1\,,\qquad \int dv\,(\mbox{non-singlet SO(5) irrep})=0\ .
\label{SO5hint}
\ee
Two basic harmonic integrals are
\be
\int dv \, v^5_a v^5_b=\frac15\delta_{ab}\,,\qquad
\int dv\, v^1_a v^2_b v^3_c v^4_d v^5_e=\frac1{5!}\varepsilon_{abcde}\ .
\label{SO5hint1}
\ee
A bit of combinatorics yields the following generalization of
these integrals~\footnote{
We (anti)symmetrize with `strength one': $[a b]=(a b-b a)/2$, $(a b)=(a b+b a)/2$, etc.}
\bea
\label{206}
\int dv\, v^5_{a_1}\ldots v^5_{a_{k}}
&=&\left\{
\begin{array}{ll}\displaystyle
\frac{3}{(2n+1)(2n+3)}\delta_{(a_1 a_2}\ldots
\delta_{a_{k-1}a_{k})}\,,\ \ & k=2n \nn\\
0\,, &k=2n+1
\end{array}
\right.
\nn\\
\int dv\, v^1_a v^2_b v^3_c v^4_d v^5_e v^5_{e_1}\ldots
v^5_{e_{k}}&=&
\left\{
\begin{array}{ll}\displaystyle
\frac{\varepsilon_{abcd(e}\delta_{e_1e_2}\ldots
\delta_{e_{k-1}e_{k})}}{8(5+2n)(2n+3)}\,,\ \ & k=2n \\
0&k=2n+1\,.
\end{array}
\right.
\eea
The gamma matrices also relate $f^{i j}$ to $X_a$,
\be
f^{ij}=\frac12\gamma^{ij}_a X_a\,,\quad
X_a=\gamma_{aij}f^{ij}\,,\quad
f^{ij}f_{ij}=-X_a X_a\,.
\ee
The sixth scalar is the $SO(5)$ singlet: $\varphi=X_6$.

\section{$\cN=2$ harmonic superspace}
\label{HSS2}

The standard $\cN=2$ harmonic superspace \cite{GIKOS,book} is based on $SU(2)/U(1)$ harmonics $u^\ui{}_i$. Here $i=1,2$ is the fundamental $SU(2)$ index and $\ui=+,-$ corresponds to the reducible representation of the $U(1)$ that defines its embedding in $SU(2)$: ${\bf 2}=(+1)+(-1)$. The harmonics $u^\ui{}_i$ form an $SU(2)$ matrix $u$ ($u u^\dagger=I_2$, $\det u=1$), so that
\bea
u^{+i}u_i^{-}=1, \quad u^{+i}u_i^{+}=u^{-i}u_i^{-}=0, \quad
u_i^{+}u_j^{-}-u_j^{+}u_i^{-}=\eps_{i j} \ ,
\eea
where the $SU(2)$ index is raised with the $SU(2)$ invariant tensor $\eps^{i j}$,
\bea
u^i=\eps^{i j}u_j, \quad u_i=\eps_{i j}u^j; \quad \eps_{i j}\eps^{j k}=\da_i^k \ .
\eea
The harmonic projections of $\cN=2$ Grassmann variables and spinor derivatives are
\bea
\ta_\al^{\pm}=u_i^{\pm}\ta_\al^i, \quad
\bar\ta_{\dot\al}^{\pm}=u_i^{\pm}\bar\ta_{\dot\al}^i, \quad
D_\al^{\pm}=u_i^{\pm}D^i_\al, \quad
\bar D_{\dot\al}^{\pm}=u_i^{\pm}\bar D_{\dot\al}^i \ .
\eea
Bosonic $SU(2)$-covariant harmonic derivatives,
\bea
D^{++}=u^{+i}\frac{\p}{\p u^{-i}}, \quad
D^{--}=u^{-i}\frac{\p}{\p u^{+i}}, \quad
D^0=u^{+i}\frac{\p}{\p u^{+i}}-u^{-i}\frac{\p}{\p u^{-i}} \ ,
\eea
form an $SU(2)$ algebra: $[D^{++},D^{--}]=D^0$, $[D^0,D^{\pm\pm}]=\pm 2D^{\pm\pm}$.
Tilde-conjugation is defined by
\bea
\wt{(u_i^{\pm})}=u^{\pm i}, \quad
\wt{(u^{\pm i})}=-u_i^{\pm}, \quad
\wt{(\ta_\al^\pm)}=\bar\ta_{\dot\al}^\pm, \quad
\wt{(\bar\ta_{\dot\al}^\pm)}=-\ta_\al^\pm \ .
\eea

In the $\cN=2$ harmonic superspace $\{x^m,\ta_\al^\ui,\bar\ta_{\dot\al}^\ui,u^\ui{}_i\}$,
there is a real analytic subspace,
\be
\{\zeta_A,u\} = \{(x_A^m,\ta_\al^{+},\bar\ta_{\dot\al}^{+}),u_i^{\pm}\}; \quad
x_A^m=x^m-i\ta^{+}\si^m\bar\ta^{-}-i\ta^{-}\si^m\bar\ta^{+}\,.
\ee
In the analytic subspace, $D^{+}$ derivatives become short,
\bea
D_{A\al}^{+}=\frac{\p}{\p\ta^{-\al}}, \qquad
\bar D_{A\dot\al}^{+}=\frac{\p}{\p\bar\ta^{-\dot\al}} \ .
\eea
Integration measures for the full superspace and its analytic subspace
are defined by
\be
\label{N2measures}
d\zeta_A^{(-4)}=d^4x_A d^4\ta^{+}\,,\quad
\int d^4\ta^{+}(\ta^{+})^2(\bar\ta^{+})^2=1\,,\quad
\int
d^8\ta(\ta^{+})^2(\ta^{-})^2(\bar\ta^{+})^2(\bar\ta^{-})^2=1\,.
\ee
The harmonic integral is defined to yield one for the singlet representation, $\int du\,
1=1$,
and zero for any other irreducible representation of $SU(2)$.
Useful examples are
\bea
\int du\, u_i^{+}u_j^{+}u_k^{-}u_l^{-} &=& \frac{1}{6}(\eps_{i k}\eps_{j l}+\eps_{i l}\eps_{j k}) \nn\\
\int du\, u_i^{+}u_j^{+}u_k^{+}u_l^{-}u_m^{-}u_n^{-} &=& \frac{1}{24}(
 \eps_{i l}\eps_{j m}\eps_{k n}
+\eps_{i l}\eps_{j n}\eps_{k m}
+\eps_{i m}\eps_{j l}\eps_{k n} \nn\\
&&\hspace{20pt}
+\eps_{i m}\eps_{j n}\eps_{k l}
+\eps_{i n}\eps_{j l}\eps_{k m}
+\eps_{i n}\eps_{j m}\eps_{k l}) \ .
\label{B9}
\eea

\subsection{$\cN=2$ gauge and hyper multiplets}

The $\cN=4$ gauge multiplet splits into one $\cN=2$ gauge multiplet and one $\cN=2$ hypermultiplet.
In $\cN=2$ harmonic superspace, these are described, respectively, by analytic prepotentials $V^{++}$ and $q^+_a=(q^+,-\bar q^+)$,
\bea
\label{analit}
D_\al^{+}V^{++}=\bar D_{\dot\al}^{+}V^{++}=0\,, \qquad
D_\al^{+}q_a^{+}=\bar D_{\dot\al}^{+}q_a^{+}=0 \ .
\eea
In addition, $\wt{V^{++}}=V^{++}$ and $\wt{q_a^{+}}\equiv q^{+a}=\eps^{a b}q_b^{+}$. The gauge supefield strengths are
\bea
W=-\frac{1}{4}(\bar{D}^{+})^2 V^{--}\,, \qquad
\bar{W}=-\frac{1}{4}(D^{+})^2 V^{--}=\wt{W} \ ,
\eea
where $V^{--}$ is uniquely defined by $D^{++}V^{--}=D^{--}V^{++}$. The superfield strengths satisfy
\be
D^{\pm\pm}W=D^{\pm\pm}\bar{W}=0\,, \quad
\bar D^\pm_{\dot\alpha} {W}= D^\pm_\alpha\bar{W}=0\,, \quad
(D^\pm)^2 W=(\bar D^\pm)^2\bar{W} \ .
\label{chir}
\ee
These constraints do not put $W$ and $q^{+}_a$ on shell. The classical action for the abelian $\cN=4$ gauge multiplet is the sum of kinetic terms,
\be
\label{clasN4}
S_{\cN=4}=\frac{1}{4}\int d^4x d^4\ta\, W^2
+\frac{1}{4}\int d^4x d^4\bar\ta\, \bar{W}^2
+\frac{1}{2}\int d^4x d^4\ta du\, q_a^{+}D^{++}q^{+a} \ .
\ee
The non-manifest $\cN=2$ supersymmetry transformations are given by
\be
\label{hN2susy}
\delta {W}=\bar\epsilon^{\dot\alpha a}\bar
D^-_{\dot\alpha}q^+_a\,, \quad
\delta \bar{W}=\epsilon^{\alpha a}D^-_\alpha
q^+_a\,, \quad
\delta q^+_a=\epsilon^\beta_a D^+_\beta{W}+\bar\epsilon^{\dot\alpha}_a
\bar D^+_{\dot\alpha}\bar{W} \ ,
\ee
where $\epsilon^{\alpha a}$ are the anticommuting parameters. Component expansions in (\ref{W-comp}) and (\ref{q-comp}) lead to canonical kinetic terms for bosonic fields.
Classical equations of motion (\ref{oncshell}) correspond to the following on-shell constraints on $W$ and $q_a^{+}$,
\be
(D^\pm)^2{W}=(\bar D^\pm)^2\bar{W}=0\,,\qquad
D^{++}q^{+a}=0\ .
\label{eom}
\ee
These follow from (\ref{clasN4}) upon varying $V^{++}$ and $q_a^{+}$, as well as from closure of (\ref{hN2susy}).

\subsection{Obtaining the WZ term by projection}
\label{B.2}

Combining the off-shell constraints (\ref{analit},\ref{chir}) and on-shell constraints (\ref{eom}), and using the algebra of covariant derivatives, we find the following on-shell identities
\bea
&&(D^-)^2 q^+_a=(\bar D^-)^2 q^+_a=0\,,\quad
(D^+)^2 q^-_a=(\bar D^+)^2 q^-_a=0\,,\nn\\&&
(D^+)^2{W}=(D^-)^2{W}=D^{+\alpha}D^-_\alpha {W}=0\,,\nn\\&&
(\bar D^+)^2\bar{W}=(\bar D^-)^2\bar {W}=\bar D^{+\dot\alpha}\bar D^-_{\dot\alpha}
\bar {W}=0\,,
\label{note2}
\eea
and
\bea
2i\partial_{\alpha\dot\alpha}q^+_a&=&\bar
D^+_{\dot\alpha}D^-_\alpha q^+_a=-D^+_\alpha \bar
D^-_{\dot\alpha}q^+_a=D^+_\alpha\bar D^+_{\dot\alpha}q^-_a
=-\bar D^+_{\dot\alpha}D^+_\alpha q^-_a\,,\nn\\
2i\partial_{\alpha\dot\alpha}q^-_a&=&
D^-_\alpha \bar D^+_{\dot\alpha}q^-_a=-\bar
D^-_{\dot\alpha}D^+_\alpha q^-_a
=\bar D^-_{\dot\alpha}D^-_\alpha q^+_a=-D^-_\alpha\bar
D^-_{\dot\alpha}q^+_a\,,\nn\\
2i\partial_{\alpha\dot\alpha}{W}&=&-\bar D^-_{\dot\alpha}D^+_\alpha{W}
=\bar D^+_{\dot\alpha} D^-_\alpha{W}\,,
\label{l128}
\eea
where $q_a^{-}=D^{--}q_a^{+}$. Noting that
\bea
\int d^8\ta\, f=\bar{D}^4 D^4 f|_{\ta=0}, \quad
\bar{D}^4 D^4=\frac{1}{2^8}
\bar D^+_{\dot\alpha}\bar D^{+\dot\alpha}
\bar D^-_{\dot\beta}\bar D^{-\dot\beta}
D^{+\alpha}D^+_\alpha D^{-\beta}D^-_\beta \ ,
\label{B18}
\eea
we calculate
\bea
\bar{D}^4 D^4 H(W,\bar{W},q_a^{+},q_a^{-})
&=&-\frac{\partial^4{H}}{\partial q^+_a \partial q^+_b \partial q^-_c \partial q^-_d}
\partial^{\alpha\dot\beta} q^-_d \partial_{\dot\alpha\alpha} q^+_c
\partial^{\beta\dot\alpha}q^+_b
\partial_{\beta\dot\beta}q^-_a
\nn\\&&
-\frac{\partial^4{H}}{\partial{W}\partial q^+_a\partial q^+_b\partial q^-_c}
\partial^{\alpha\dot\beta}{W}\partial_{\alpha\dot\alpha}
q^+_c \partial^{\beta\dot\alpha}q^+_b
\partial_{\beta\dot\beta}q^-_a
\nn\\&&
-\frac{\partial^4{H}}{\partial{W}\partial q^+_a\partial q^-_c \partial q^-_d}
\partial^{\alpha\dot\beta}q^-_d \partial_{\alpha\dot\alpha}q^+_c
\partial^{\beta\dot\alpha}{W}
\partial_{\beta\dot\beta}q^-_a+\ldots
\,, \qquad
\label{l130}
\eea
where we have shown only terms with cyclic contraction of the spinor indices. Upon projecting to $\ta=0$, we find the expression for $\Ga_{WZ}$ given in (\ref{l135}). Alternatively, one could use $\ta$-expansions (\ref{W-comp}) and (\ref{q-comp}) and integrate by the rule given in (\ref{N2measures}).



\end{document}